\documentclass[12pt,a4paper]{article}

\usepackage{amsmath,amssymb,amsfonts,bm,graphicx,subfiles,xspace}
\usepackage{tikz}

\newlength{\abstwidth}
\def\epp{$e^+e^-$\xspace}
\def\ep{$e$p\xspace}
\def\pp{pp\xspace}
\def\pA{p$A$\xspace}
\def\AA{$AA$\xspace}
\def\pythia{\textsc{Pythia}~}
\def\herwig{\textsc{Herwig}~}
\def\sherpa{\textsc{Sherpa}~}

\newcommand{\figref}[1]{fig.~\ref{#1}\xspace}
\newcommand{\secref}[1]{sec.~\ref{#1}\xspace}
\renewcommand{\eqref}[1]{eq.~(\ref{#1})\xspace}

\begin{document}
\sloppy
 
\pagestyle{empty}
 
\begin{flushright}
MCnet-24-01\\
\end{flushright}

\vspace{\fill}

\begin{center}
{\Huge\bf String interactions as a source of collective behaviour}\\[4mm]
{\Large Christian Bierlich} \\[3mm]
{\texttt christian.bierlich@fysik.lu.se}\\[1mm]
{\it Division of Particle and Nuclear Physics,}\\[1mm]
{\it Department of Physics, Lund University}\\[1mm]
\end{center}

\vspace{\fill}

\begin{center}
\begin{minipage}{\abstwidth}
{\bf Abstract}\\[2mm]
The discovery of collective effects in small collision systems, have spurred a renewed interest in hadronization models, also as a source for collective effects all the way to large collision systems, where they are usually ascribed to the creation of a Quark--Gluon Plasma. In this topical mini-review, the microscopic model for string interactions, based on the Lund string hadronization model, developed with exactly this aim is reviewed, and some prospects for the future presented.
\end{minipage}
\end{center}

\vspace{\fill}

\phantom{dummy}

\clearpage

\pagestyle{plain}
\setcounter{page}{1}

\section{Introduction}

Collisions of heavy nuclei at high energy are most often analyzed under the assumption that a Quark-Gluon Plasma (QGP) is created fractions of a fm after the collision. As evidence for QGP creation, several experimental signatures are considered. These include observations\footnote{QGP formation predicts several other types of modifications with respect to proton collisions, such as jet quenching, heavy quark modifications, altered resonance production, and several more. The focus in this mini-review will, however, be on strangeness and flow.} of strangeness enhancement relative to the proton-proton baseline \cite{Rafelski:1982pu,NA35:1991ece} and anisotropic flow \cite{Ollitrault:1992bk,STAR:2000ekf}. Over the past decades, a rich research programme has emerged, studying the properties of the QGP.

The QGP paradigm differs vastly from the dynamics traditionally assumed to govern high-energy collisions of protons. In the latter, models based on the formation of strings \cite{Andersson:1983ia} or clusters \cite{Webber:1983if}, which subsequently decay into hadrons, have dominated the field. These models are implemented in the highly successful ``General purpose Monte Carlo event generators'' \cite{Buckley:2011ms,Campbell:2022qmc}, such as \pythia \cite{Bierlich:2022pfr}, \herwig \cite{Bellm:2015jjp}, and \sherpa \cite{Sherpa:2019gpd}. In proton collisions, this includes the scattering of multiple partons in each collision event, which, in the case of strings, leads to multiple strings connecting the emerging partons with each other and the beam remnants. No formation of QGP has traditionally been assumed, and the models have done remarkably well in reproducing most features of \epp, \ep, and \pp collisions.

Over the past decade, the distinction between small (\epp, \ep, and \pp) and large (\pA and \AA) collision systems has become much less clear. Measurements from the LHC have shown that \pp collisions exhibit similar qualitative features of strangeness enhancement \cite{ALICE:2016fzo} and flow \cite{CMS:2010ifv}, as observed in larger collision systems (as well as many more recent observations. See ref. \cite{ALICE:2022wpn} for a recent review of ALICE results). There are even indications that flow signals can potentially form in \epp, as shown by re-analyzed data from ALEPH \cite{Chen:2023nsi}. This raises the question of whether a QGP is also created in small collision systems, challenging the ``general purpose'' paradigm of the aforementioned event generators. This assumption is successfully pursued in, for example, core-corona models \cite{Werner:2007bf,Kanakubo:2019ogh}. Another possibility is to consider whether dynamics not including QGP formation could be responsible for the same signatures. This latter possibility is the underlying assumption of the microscopic model of interacting Lund strings, which has been developed over the past years, and is the topic of this mini-review\footnote{This review is thus \emph{not} a general review on collectivity in small systems. The reader is referred to excellent reviews in refs. \cite{Li:2012hc,Schlichting:2016sqo,Nagle:2018nvi,Citron:2018lsq,Adolfsson:2020dhm,ALICE:2022wpn} for other perspectives, both from an experimental and theoretical starting point. In particular ref. \cite{Nagle:2018nvi} offers a RHIC perspective, which is not covered in this review.}.

The main aim of the model is to establish a QGP-free baseline for predictions and postdictions for experimental measurements in heavy ion collisions. 

The mini-review is structured as follows. In \secref{sec:mpi} the underlying Angantyr model is briefly described, before diving into the string formalism in \secref{sec:string}. Here interactions of multi-string systems is treated through \secref{sec:multi-string}, before some concluding remarks and an outlook is given in \secref{sec:conclusion}.

\begin{figure}
         \begin{tikzpicture}[every node/.style={scale=0.7}, align=left]
                 \draw [->] (0.5,0) node[below]{Projectile} -> (5.5,0) node[below]{Target~~\scalebox{1.5}{$\eta$}};
                 \draw [->] (3,0) -> (3,3) node[left]{\scalebox{1.5}{$\frac{\mathrm{d}N}{\mathrm{d}\eta}$}};
                 \draw (1,0) -> (5,0.5);
                 \draw (1, 0.5) -> (5,0);
                 \draw [red,dashed] (1, 0.5) -> (5, 0.5) node[right]{pp\\ collision};
                 \draw (1, 0.) -> (5, 1.0);
                 \draw (1, 0.) -> (5, 1.5);
                 \draw (1, 0.) -> (5, 2.0);
                 \draw (1, 0.) -> (5, 2.5);
                 \draw [red,dashed] (1, 0.5) -> (5, 2.5) node[right]{p$A$\\ collision};
                 \draw (1, 1.0) -> (5, 0.);
                 \draw (1, 1.5) -> (5, 0.);
                 \draw (1, 2.0) -> (5, 0.);
                 \draw (1, 2.5) -> (5, 0.);
                 \draw [red,dashed] (1, 2.5) node[above]{$AA$\\ collision} -> (5, 2.5);
        \end{tikzpicture}
	\includegraphics[width=0.5\textwidth]{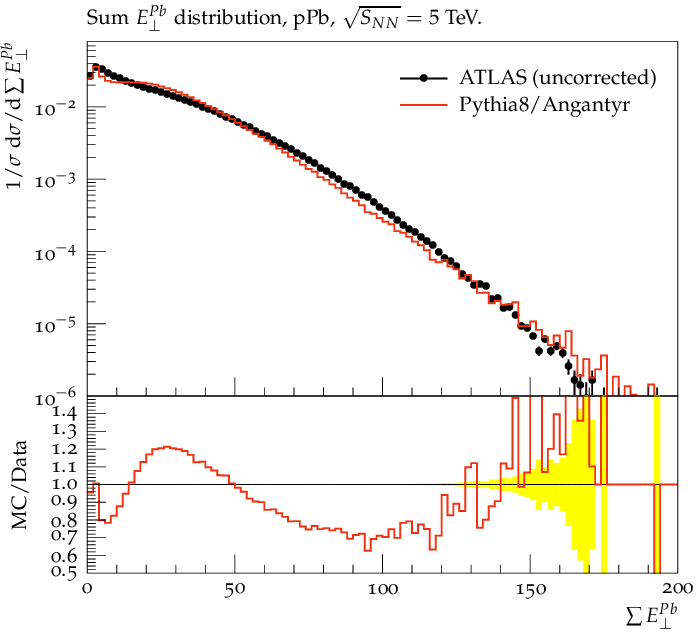}
\caption{\label{fig:angantyr-mult-1}The main concept of Angantyr (left) is to build up a heavy ion collision from contributions from each wounded nucleon in the projectile and target. The figure shows how this can be used to build a \pp, a \pA and an \AA collision. This treatment, along with emphasis on colour fluctuations in the initial state, provides a good description of forward production (right), important to determine centrality classes of collisions in the same way as experiments does. In this figure, and throughout the paper, data comparison is done using RIVET \cite{Bierlich:2019rhm,Bierlich:2020wms}. (Right figure from ref. \cite{Bierlich:2018xfw}, data from ATLAS \cite{ATLAS:2015hkr})}
\end{figure}
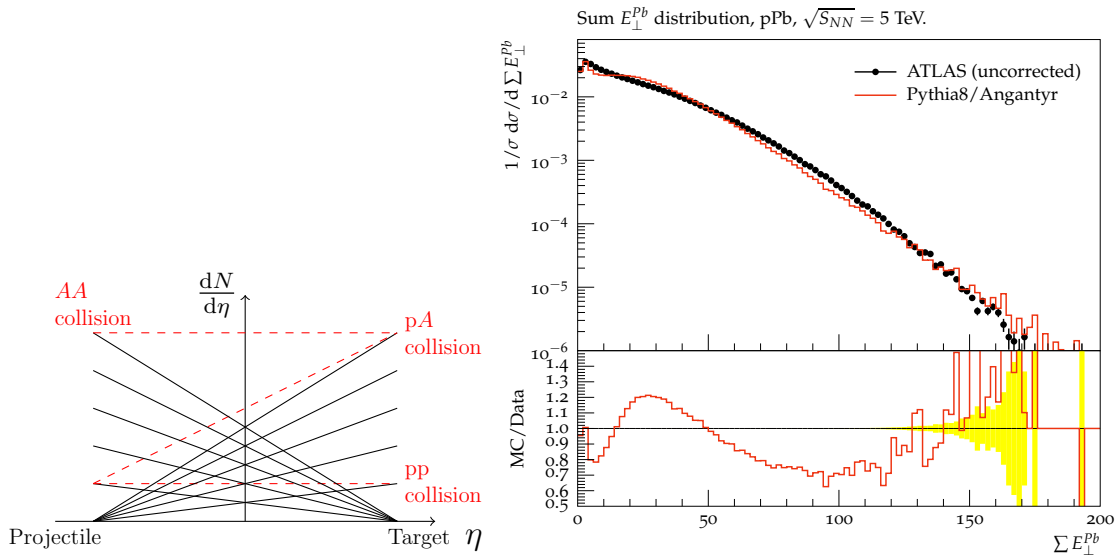

\section{Multi-parton interactions and the Angantyr model}
\label{sec:mpi}

Strings are produced by partons created in hard collisions or subsequently radiated off in the parton shower. For multi-parton interactions (MPIs) in proton collisions, the main model by Sjöstrand and van Zijl \cite{Sjostrand:1987su} treats all parton-parton scatterings as independent $2\rightarrow 2$ processes, with a scattering cross section calculable in perturbative QCD, which has the approximate behaviour \cite{Bierlich:2022pfr}:
\begin{equation}
\label{eq:mpi-pt}
		\frac{\mathrm{d}\hat{\sigma}}{\mathrm{d}p^2_\perp} \propto \frac{\alpha^2_s(p^2_\perp)}{p^4_\perp}.
\end{equation}
The expression is evidently divergent as $p_\perp \rightarrow 0$, and as such it will at some point exceed the total nucleon-nucleon cross section, which is clearly unphysical. Even with a cut-off scale of 1 GeV, a reasonable value for when one would expect perturbative QCD to break down, is too small. Instead, a screening scale $p_{\perp,0}$ is introduced, entering as a dampening factor, modifying \eqref{eq:mpi-pt} as:
\begin{equation}
\label{eq:mpi-pt-mod}
		\frac{\alpha^2_s(p^2_\perp)}{p^4_\perp} \mapsto \frac{\alpha^2_s(p^2_\perp + p^2_{\perp,0})}{(p^2_\perp + p^2_{0,\perp})^2}. 
\end{equation}
This cut-off is usually interpreted as colour screening scale, though suggestions linking it to saturation in the spirit of Color Glass Condensate (CGC) \cite{Gelis:2010nm} instead, have been put forth. 

The extension of this model, Angantyr \cite{Bierlich:2016smv,Bierlich:2018xfw}, underpins all heavy ion results presented in this review. Before it is briefly outlined in the following, it should be mentioned that it is not at all obvious why a model assuming that soft particle production can be correctly described with $2\rightarrow 2$ parton scatterings as a starting point. Other models, such as the already mentioned CGC models, as well as the also already mentioned EPOS model, have other starting points. In the latter case, soft inelastic processes are represented directly as cut Pomerons.

\subsection{Angantyr for heavy ion collisions}

Angantyr is initialized by a Glauber calculation \cite{Miller:2007ri,Rybczynski:2013yba} with particular emphasis on colour fluctuations in the initial state \cite{Gribov:1968jf,Heiselberg:1991is,Blaettel:1993ah}. Through the Good-Walker formalism \cite{Good:1960ba} these can be translated to semi-inclusive nucleon-nucleon cross sections, which can be used to fit the parameters of the initial state model from \pp data only. Crucially, this treatment can distinguish not only which nucleons are hit, and which are spectating, but also \emph{how} the nucleons are hit, i.e. whether they participate elastically, diffractively or absorbed through a colour exchange.

Once it is determined which nucleons interact, multiple partons from each nucleon can interact, in the same spirit as the MPI model for \pp. The \textit{ansatz} for the Angantyr model, however, is the wounded nucleon model\footnote{In this sense, Angantyr is more of a successor to the FRITIOF model \cite{Andersson:1986gw}, which had the same starting point.} \cite{Bialas:1976ed}. In this model, an \textit{ad hoc} multiplicity function is applied to each wounded nucleon in the collision, which can then be stacked to construct a \pA or an \AA collision. The process is sketched in \figref{fig:angantyr-mult-1} (left). Instead of a function, Angantyr uses a dynamical model, where a wounded nucleon is generated as normal MPIs, but with reduced phase space. Two wounded nucleons correspond to a \pp collision, and parameters of the model can, therefore, be fitted to \pp only. Interestingly, it is possible to interpret the addition of an additional wounded nucleon to a \pp collision (making it a proton-Deuteron collision) in the cut Pomeron language \cite{Bierlich:2018xfw}. This may point to the reasons why the seemingly very different models such as the CGC and EPOS models mentioned above, and the Angantyr model, can reproduce very similar physics, from a very different starting point. This is a direction of research currently left very unexplored.

As shown in \figref{fig:angantyr-mult-1} (right), this procedure reproduces centrality measures (here $\sum E_\perp$ in the forward direction, which is the ATLAS choice of centrality measure) in \pA well (the model performs similarly well for \AA \cite{Bierlich:2018xfw}). This is essential for providing direct comparisons to centrality binned experimental data, where centrality is experimentally determined by cutting percentiles in this quantity.

\begin{figure}
	\includegraphics[width=0.5\textwidth]{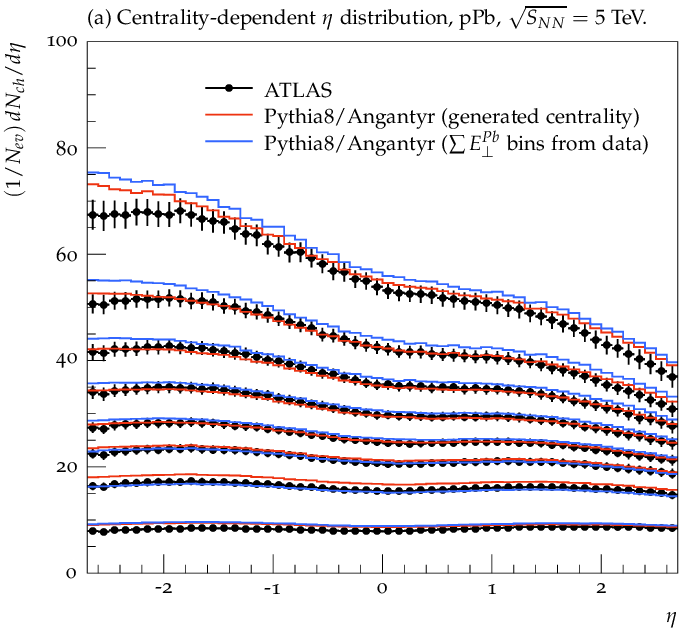}
	\includegraphics[width=0.5\textwidth]{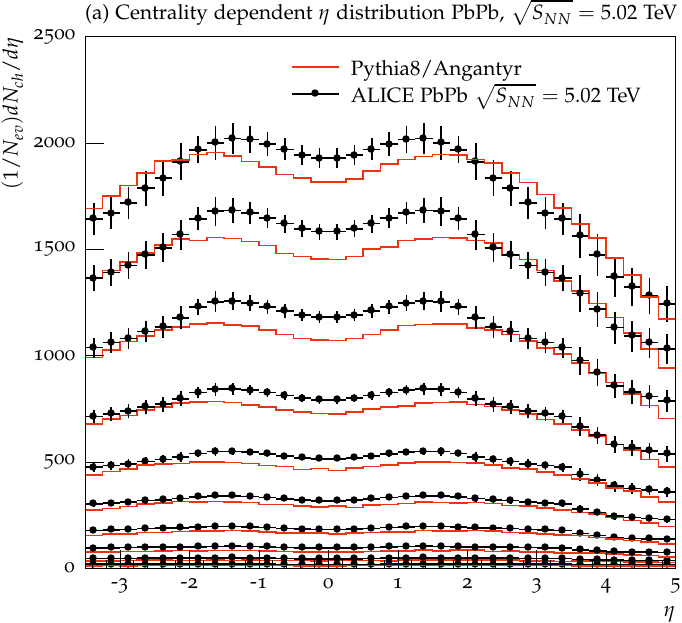}
\caption{\label{fig:angantyr-mult-2}Particle production at mid-rapidity from the Angantyr model, compared to data in pPb collisions (left) and PbPb (right) in centrality bins. (Figures from ref. \cite{Bierlich:2018xfw}, data from ATLAS and ALICE \cite{ATLAS:2015hkr,ALICE:2016fbt})}
\end{figure}

Relating the forward production (centrality bins) to mid-rapidity production yields the comparisons shown for pPb in \figref{fig:angantyr-mult-2} (left), and for PbPb in \figref{fig:angantyr-mult-2} (right). This basic extension of the MPI model to heavy ion collisions performs very well for total multiplicities, as observed.

The primary purpose of the Angantyr model is to provide a ``blank canvas'' upon which microscopic, string-based models for collectivity can be constructed. This is the main topic of this mini-review, which will be expanded upon in the coming sections.

\section{The Lund string model and microscopic collectivity}
\label{sec:string}

The basis of the microscopic model for collectivity lies in the so-called Lund string. This is the physics model underlying string fragmentation in the \pythia Monte Carlo event generator. It represents a specific version of a more general class of ``hadronic string'' or ``flux-tube'' models, all of which make distinct phenomenological assumptions about the dynamics, particularly of string breakups. In this section, the main components\footnote{For a more comprehensive, recent review, see ref. \cite{Bierlich:2022pfr}.} of the Lund string model for a single string will be introduced in \secref{sec:single-string}, proceeding to systems of multiple strings in \secref{sec:multi-string}. Colour reconnection is presented first in \secref{sec:colour-reconnection}, with string shoving and rope hadronization presented in \secref{sec:string-shoving} and \ref{sec:rope-hadronization}, respectively.

\begin{figure}
	\includegraphics[width=0.5\textwidth]{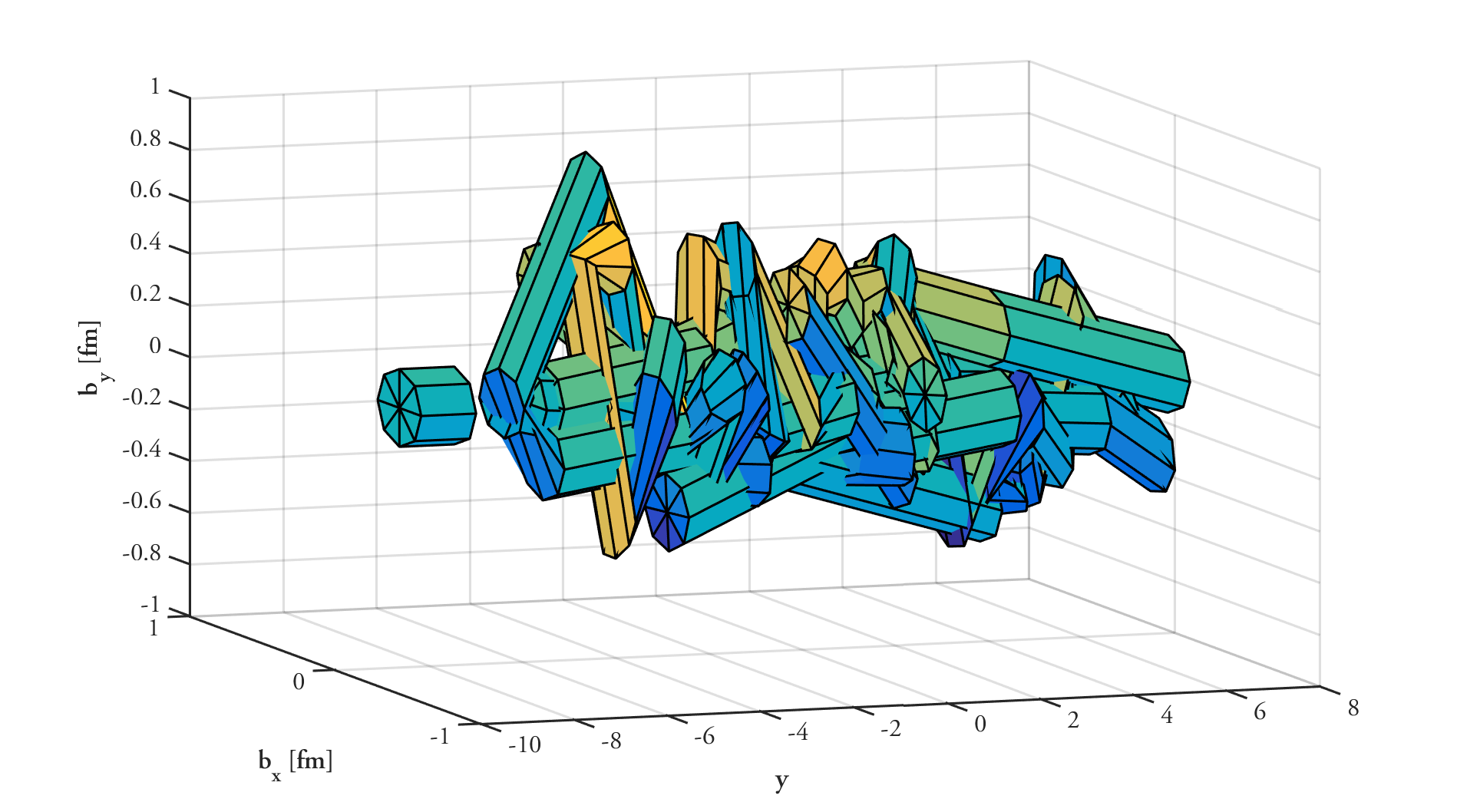}
	\includegraphics[width=0.5\textwidth]{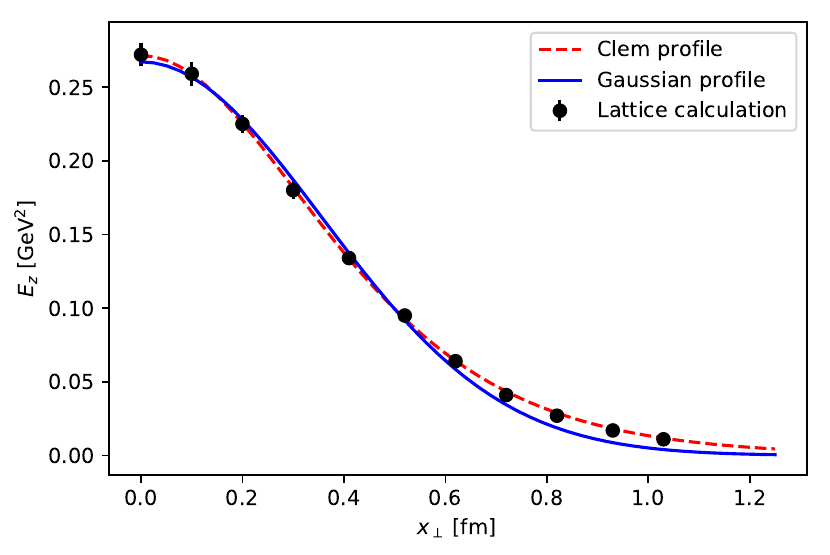}
\caption{\label{fig:many-strings}In pp collisions at $\sqrt{s} =$7 TeV (left), many strings can overlap in transverse space and rapidity as shown, giving rise to interactions and coherence effects. Note that the string radius is here put at 0.2 fm for illustrative purposes, but can in reality be much larger. The string's transverse (colour)-electric field itself is not a cylinder, but can rather be approximated as a Gaussian (right), with black points being lattice calculations \cite{Cea:2014uja}. (Figures from refs. \cite{Bierlich:2014xba,Bierlich:2020naj})}
\end{figure}

\subsection{Physics of a single Lund string}
\label{sec:single-string}

The Lund string model is built upon the observation from lattice QCD (and previously from onium spectra) that the confining potential between a $q\bar{q}$ pair is linear, possessing a string tension of approximately $\kappa \approx$ 1 GeV/fm. This is given a physical interpretation as flux tubes, akin to vortex lines in a superconductor. The Lund string model \cite{Andersson:1983ia} is a 1+1 dimensional phenomenological model realization of this concept. It is based on the assumption that the fragmentation dynamics are insensitive to the width of the string, allowing it to be approximated by a massless relativistic string of infinitesimal width. Mesons are modelled as small $q\bar{q}$ string pieces (``yo-yo'' modes), produced by successively fragmenting pieces off the original string. The main components of the original string model include the breakup mechanism dictating the phase space distribution of produced hadrons, as well as the inclusion of gluons as kinks on the string.

The string breaking can be modelled as an iterative process, where each break produces a new hadron. With minimal constraints, amounting to the requirement that the same result should be obtained whether fragmenting from one end of the string or the other, the distribution of the momentum fraction ($z$) of the remaining light-cone momentum taken away by the hadron produced in each splitting is given as:
\begin{equation}
\label{eq:frag-fun}
	f(z) \propto \frac{(1-z)^a}{z}\exp{\left(-\frac{b m^2_\perp}{z}\right)},
\end{equation}
where $m_\perp$ is the hadron's transverse mass ($m^2_\perp = m^2 + p^2_\perp$), and $a$ and $b$ are parameters determined from experiments. This formulation fixes the longitudinal degrees of freedom in the string breaking process. The transverse momentum and flavour/mass of the quarks are determined from the probability of the $q\bar{q}$ pair tunnelling through a region of size $m_\perp/\kappa$, given by:
\begin{equation}
\label{eq:mpt}
	\frac{\mathrm{d}\mathcal{P}}{\mathrm{d}^2p_\perp} \propto \kappa \exp(-\pi m^2_{\perp,q}/\kappa) = \exp(-\pi m^2_q/\kappa) \times \exp(-\pi p^2_{\perp,q}/\kappa).
\end{equation}
Due to the factorization in the above expression, the transverse momenta and mass of the quarks can be treated separately. In practice, this is done by generating Gaussian $p_\perp$-kicks, with a parameter $\sigma_{p_\perp}$ governing the width of the distribution\footnote{In principle, $\sigma_{p_\perp} = \kappa/\pi \approx (0.25$ GeV$)^2$, but fits to LEP data suggest this number to be higher, indicating that a non-negligible factor of the $p_\perp$ comes from another source.}. The suppression of strange quarks relative to u or d quarks is described by a factor $\rho = \exp(-\pi(m^2_s - m^2_u)/\kappa)$, which is also treated as a parameter\footnote{In principle, this number could be obtained if a suitable value for quark masses were inserted. However, current quark masses lead to too little strangeness suppression, and constituent quark masses to too much.}.

\subsection{Interactions of multiple strings}
\label{sec:multi-string}

In collisions of protons (and thus also in larger collision systems), strings will overlap with each other (see \figref{fig:many-strings}, left). It is therefore reasonable to question whether the initial assumption of strings being infinitely thin should be relaxed. Discarding the assumption alltogether, allowing strings to have a transverse extension, naturally leads to string interactions. Results from lattice QCD \cite{Cea:2014uja} can estimate the field shape, as shown in the black points in \figref{fig:many-strings}. As indicated by the fitted curve, the field is well approximated by a Gaussian\footnote{The other curve, labelled "Clem profile," is derived from Landau-Ginzburg theory and is given by $E = K_0(\sqrt{\rho^2 + \xi^2_v}/\lambda)$ (where $K_0$ is a modified Bessel function, $\rho$ the cylindrical radius coordinate, $\xi_v$ the condensate coherence length, and $\lambda$ the penetration depth.) See ref. \cite{Bierlich:2020naj} for more details.}. However, the transverse size of the string's electrical field as shown in the figure should be taken with caution. Since lattice calculations are performed in arbitrary ``lattice units'', a connection to a physical quantity is necessary to translate to physical units. Typically, the string tension is used for such calculations, but further assumptions are needed about the fraction of energy going into the field versus the fraction needed to break the condensate. Therefore, these types of calculations cannot provide a precise estimate of the string width, but they do give a good idea of the order — which is significant enough for string interactions to occur.

In the Lund string model, the string is produced by chaining together quarks and gluons, which were previously produced by the hard scattering and successive emissions from the parton shower. Due to causality, the field is initially very thin, and interactions between strings do not depend on overlapping fields. Instead, they are formulated in terms of ``colour reconnections'', the term given to sub-leading colour corrections to the string topology provided by the parton shower. It is only after this phase that the string reaches a size where the (colour)-electric fields start interacting as fields, pushing each other apart. Eventually, the strings will hadronize, and if they are still overlapping, they may hadronize with a larger string tension than $\kappa = 1$ GeV/fm. After hadronization, hadrons may interact with each other in a hadronic cascade.

Importantly, the time allowed for the strings to interact with each other is \emph{not} a free parameter of the model. The derivation of the fragmentation function \eqref{eq:frag-fun} also yields, as a by-product, the distribution of string breakup vertices in proper time ($\tau$). The distribution is more conveniently written in terms of $\Gamma = (\kappa \tau)^2$, given by:
\begin{equation}
	\mathcal{P}(\Gamma)\mathrm{d}\Gamma = \Gamma^a \exp(-b \Gamma) \mathrm{d}\Gamma.
\end{equation}
From this distribution, the average string break time is:
\begin{equation}
	\langle \tau^2 \rangle = \frac{1+a}{b\kappa^2},
\end{equation}
which, with reasonable values for $a$ and $b$ (determined from total multiplicities in \epp using \eqref{eq:frag-fun}), gives $\langle \tau^2 \rangle \approx 2$ fm, albeit with a tail to larger times. While this cannot be directly interpreted as \emph{the} hadron formation time — it is unclear from the model whether a hadron can be considered produced once the string breaks or after, for example, one full oscillation of the yo-yo mode — this figure establishes some conceptual facts about the microscopic model. First and foremost, the time allowed for strings to interact before hadronization is significantly shorter (a factor of 5-10) than in hydrodynamic models assuming a QGP. Secondly, there is no single time defined for ``string freezeout''. The process will inevitably result in some strings hadronizing earlier and some later\footnote{This, in principle, ought to give rise to a mixed phase of strings and hadrons. However, this possibility has not yet been pursued in this framework.}. Third, the earlier hadronization compared to QGP-based models allows the subsequent hadronic cascade to have a much larger impact on observable results. This latter effect is not discussed in this review, but interested readers are referred to refs. \cite{daSilva:2020cyn,Bierlich:2021poz}.

Ordered roughly by time after the initial collision, the microscopic model for collective behaviour consists of the following components:
\begin{itemize}
\item $\tau \approx$ 0-0.6 fm: Parton shower and colour reconnection. Emissions down to the parton shower cut-off scale and altered colour topologies, which can give rise to short-range collective effects \cite{OrtizVelasquez:2013ofg,Bierlich:2018lbp} and enhanced baryon production \cite{Christiansen:2015yqa,Bierlich:2015rha}.
\item $\tau \approx$ 0.6-1.4 fm: String shoving. Strings reach a transverse size where they interact maximally before hadronizing, giving rise to soft collective effects \cite{Bierlich:2017vhg}, and affecting jets as well \cite{Bierlich:2019ixq}.
\item $\tau \approx$ 1.4 fm: Rope hadronization. Strings can hadronize in a colour multiplet if they are still sufficiently close together, forming a so-called colour rope \cite{Biro:1984cf}, which leads to enhanced strangeness and baryon production \cite{Bierlich:2014xba}.
\item After $\tau \approx$ 1.4 fm: Hadronic cascade. Hadronic rescattering \cite{Sjostrand:2020gyg} results in significant corrections to both flavour and flow, particularly in the dense environment of \AA collisions \cite{Bierlich:2021poz}.
\end{itemize}

\subsection{Colour reconnections}
\label{sec:colour-reconnection}

The colour topology delivered by the matrix element calculation and parton shower, uses the ``leading colour'' limit\footnote{Formally obtained \cite{tHooft:1973alw} by letting $N_c \rightarrow \infty$ ($N_c$ is the number of colours), while keeping $\alpha_s N_c$ fixed. This eliminates coherence effects which are suppressed by a factor $1/N^2_c$, and allows for a simple representation of gluons as colour-anti-colour.}, in which the string topology is uniquely determined by the fact that the probability of two random colours are the same, vanishes. As an example, two $q\bar{q}$ pairs in the leading colour limit will never be allowed to form a singlet, but will always form an octet. Thus in $\mathbf{3} \otimes \mathbf{\bar{3}} = \mathbf{8} \oplus \mathbf{1}$, the $\mathbf{1}$ vanishes in the limit.

\begin{figure}
	\includegraphics[width=0.5\textwidth]{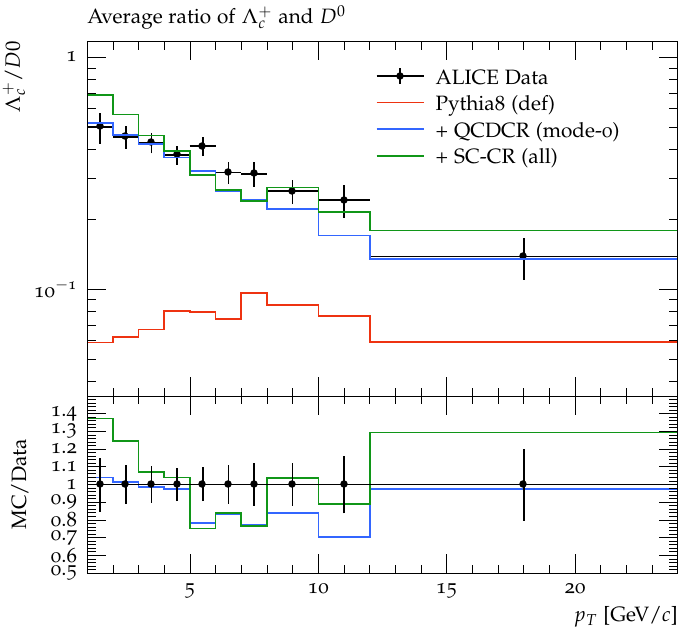}
	\includegraphics[width=0.5\textwidth]{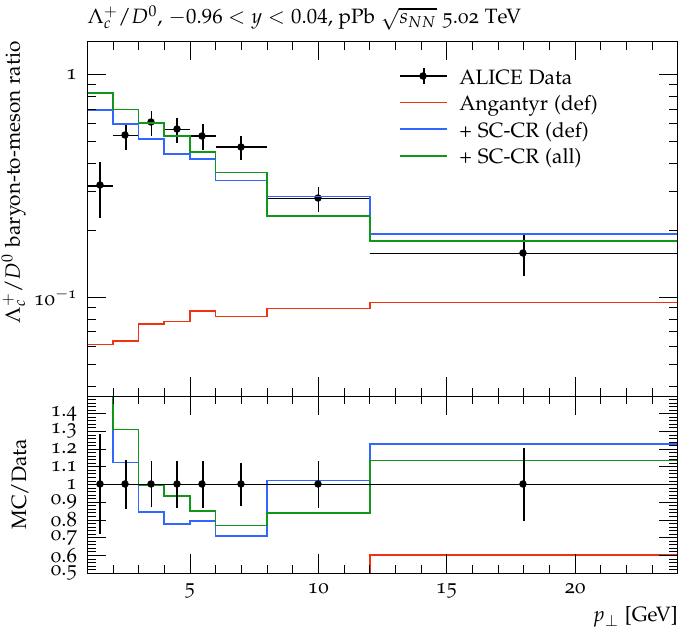}
\caption{\label{fig:cr-results}Results from QCD-CR colour reconnection in pp (left) and pPb (right), showing that the junction production mechanism drastically improves the description of $\Lambda_c/D$ ratios in both collision systems.(Figures from ref. \cite{Bierlich:2023okq}, data from ALICE \cite{ALICE:2020wfu,ALICE:2021rzj})}
\end{figure}

Colour reconnection models are an attempt to reintroduce the colour topologies which \emph{should} have been generated, had the perturbative calculation been done in full colour. The simplest reconnection models aims at reintroducing the singlet. This is for example the case for the default \pythia~reconnection model \cite{Sjostrand:1987su} (often referred to as ``the MPI based model''). The main motivation for the model, was the observation that with completely independent MPIs, $\langle p_\perp \rangle$ will not rise with multiplicity, as was observed at UA1 \cite{UA1:1989bou}. If MPIs are allowed to reconnect, and in essence transfer some multiplicity to $p_\perp$, this is achieved. 

Since one cannot perform a full-colour correction from first principles, colour reconnection models contain some \textit{ad hoc} modelling element. In the case of string based models, this is often centred around the notion of reducing total string length. This is quantified in the so-called $\lambda$-measure, which measures how many hadrons of some reference mass $m_0$, a given string piece has room (in phase space) to produce. For a simple $q\bar{q}$ system it can easily be written as e.g. $\lambda_{q\bar{q}} = \ln(m^2_{q\bar{q}}/m^2_0)$, but for more complicated topologies approximate expressions must be used, as e.g.:
\begin{equation}
	\lambda = \sum^n_{i=0} \ln\left(1 + \frac{(k_ip_i + k_{i+1}p_{i+1})^2}{m^2_0} \right),
\end{equation}
where $k$ is 1 for quarks and $1/2$ for gluons, as the gluon momentum is shared between the two connecting string pieces.

It has been shown that even the simplest colour reconnection model in \pythia can give rise to some QGP-like effects, in particular baryon-meson ratios and flow \cite{OrtizVelasquez:2013ofg}. The effects are, however, only short range in rapidity \cite{Bierlich:2018lbp} and can therefore not explain observations such as the pp ridge \cite{CMS:2010ifv} or $v_2$ with rapidity gap between particles of interest.

More advanced colour reconnection models have been developed, in particular to further address baryon-meson ratios. Most noteworthy is the so-called sub-leading colour or ``QCD-CR'' model \cite{Christiansen:2015yqa}. The model aims to address also the higher neglected multiplets when constructing string topologies, rather than just the singlet, for example the anti-triplet in $\mathbf{3} \otimes \mathbf{3} = \mathbf{6} \oplus \mathbf{\bar{3}}$, or the triplet and anti-sextet in $\mathbf{3} \otimes \mathbf{8} = \mathbf{15} \oplus \mathbf{\bar{6}} \oplus \mathbf{3}$. Importantly, the $\mathbf{\bar{3}}$ and $\mathbf{\bar{6}}$ in these states, involves colour states\footnote{Such states have no analogy in $N_c \rightarrow \infty$ limit, which corresponds only to dipole-like connections. These states represent $\epsilon_{ijk}$ structures in colour space, which are explicitly $N_c = 3$.} which in the string model are interpreted as ``junctions'': states where three string legs are connected to a single point, and which carry intrinsic baryon number \cite{Sjostrand:2002ip}. Allowing the formation of such states in colour reconnection (in pairs of junction-anti-junction, so it does not violate baryon number conservation) increases the amount of produced baryons, when a high number of MPIs increases the chance of a junction colour reconnection. Technically, one performs an approximate calculation of the weight factors based on the $\mathbf{SU(3)}$ algebra, determining which configurations are more or less likely. The actual configuration is then chosen based on which of the possible configurations minimizes the $\lambda$-measure the most.

The model was introduced to address discrepancies of baryon yields in the first years of LHC running wrt. \epp (see e.g. ref. \cite{CMS:2011jlm}), but crucially also allows for production of double heavy flavour baryons, not allowed in the normal string model. In particular production of heavy flavour baryons have been a large success for the model, where the $\Lambda_c/D$ ratio vs $p_\perp$ in pp collisions (see \figref{fig:cr-results} (left)) was found by ALICE \cite{ALICE:2020wfu} to differ drastically from expectation. The colour reconnection model gives a satisfactory description. In ion collisions later refinements of the model \cite{Bierlich:2023okq}, taking into account the non-negligible mass of the charm quark, allowed predictions also for \pA, as shown in \figref{fig:cr-results} (right). Also here the description is drastically improved.

The last word is, however, not said for this type of models. Further studies of charm baryon yields have revealed that for example charmed $\Xi$ is not equally well reproduced \cite{Bierlich:2023okq,ALICE:2021psx}. Some is on the account of further enhancement of strangeness (see \secref{sec:rope-hadronization}), but not all can be explained this way. Colour reconnection remains to be an active area of study.
 
\subsection{The string shoving mechanism}
\label{sec:string-shoving}

As explained in the beginning of \secref{sec:multi-string}, the transverse (colour-)electric field can be approximated well as a Gaussian. After the string has had time after its initial creation to expand to its full transverse size, strings will start ``shoving'' each other, with a force calculable from the field. If the field is $E = N\exp(-\rho^2/2R^2)$, where $\rho$ is the radius in cylindrical coordinates and $R$ is the equilibrium radius. $N$ is a normalization factor, which is determined by letting the energy in the field correspond to a fraction $g$ of the total string tension.

It is now assumed that all multi-string colour configurations such as $\mathbf{3} \otimes \mathbf{3}$ have already been correctly colour reconnected. This means that only octet configurations are left, while singlets have vanished. Remaining string dipoles will therefore only repel each other, never attract. The energy per unit length of two strings overlapping is $\int \mathrm{d}^2\rho (\mathbf{E}_1 + \mathbf{E}_2)^2/2$, from which the force between two strings transversely separated by $d_\perp$ can be calculated as:
\begin{equation}
\label{eq:shoving}
	f(d_\perp) = \frac{g\kappa d_\perp}{R^2} \exp\left(-\frac{d^2_\perp}{4R^2}\right).
\end{equation}

This is the main ingredient of the string shoving mechanism. Technicals complications arise when implementing this in a Monte Carlo. String pieces are not aligned in parallel, but a suitable Lorentz transformation can be found to align each pair in parallel planes \cite{Bierlich:2020naj}. Also, simply doing a step-wise time evolution, is not practical from a computational point of view. Instead the shoving model is implemented in similar spirit as a parton shower, ordered in the $p_\perp$ pushes.

\subsubsection{Soft collective effects}

\begin{figure}
	\includegraphics[width=0.5\textwidth]{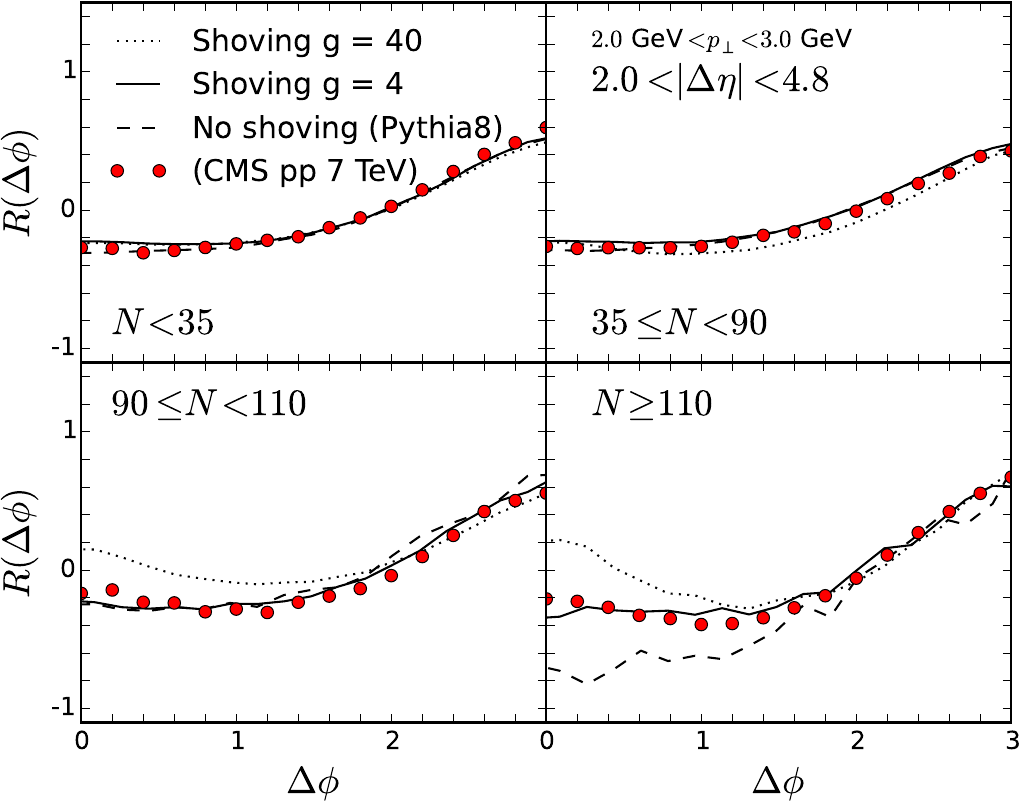}
	\includegraphics[width=0.5\textwidth]{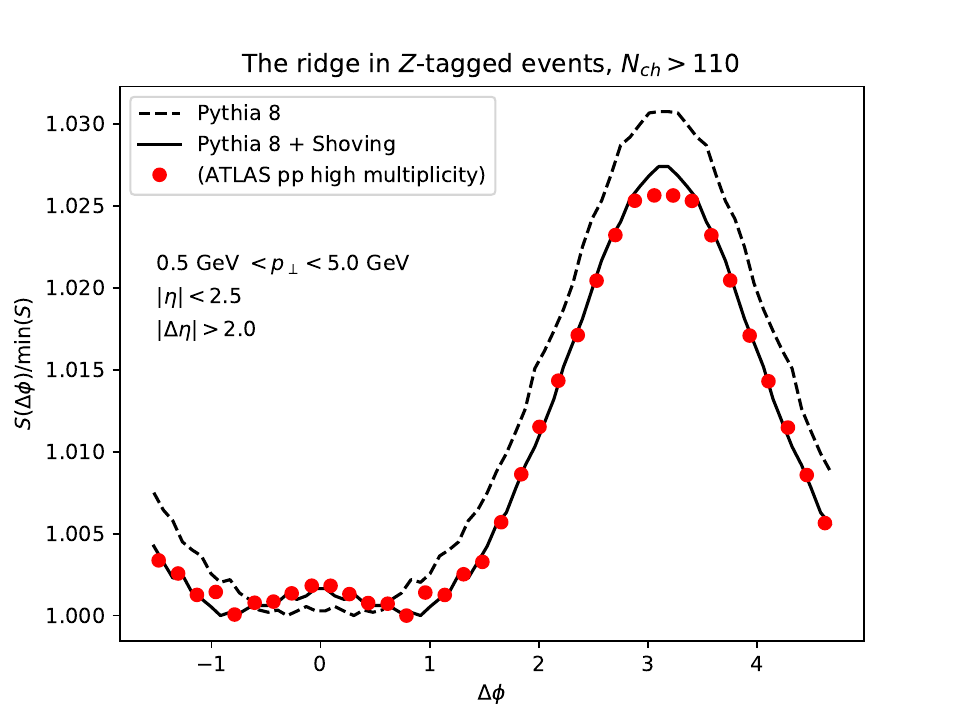}
\caption{\label{fig:shoving-results-pp}The string shoving model reproducing the ridge in pp collisions in minimum bias collisions in different multiplicity bins (left), and in $Z$-tagged, high multiplicity collisions (right). (Figures from ref. \cite{Bierlich:2017vhg,Bierlich:2019ixq}, data from CMS and ATLAS \cite{CMS:2010ifv,ATLAS:2019wzn}.)}
\end{figure}

The main intent of the shoving model, is to be able to generate anisotropic flow as a response to the spatial initial conditions. In proton collisions the initial conditions are currently just produced as the convolution of two Gaussian mass distributions, which is itself a Gaussian, i.e. fully symmetric. Higher order flow is therefore produced only from fluctuations in the initial conditions, and not from intrinsic asymmetry\footnote{It is possible to calculate initial geometries in other types of approaches like BFKL evolution in impact parameter space \cite{Bierlich:2019wld}. While first steps have been made to interface this, it is not yet mature enough for real calculations.}. 
Overall observables, such as the long-range ridge first measured in pp by CMS \cite{CMS:2010ifv}, is well reproduced as shown in \figref{fig:shoving-results-pp}. (Note that the parameter $g$ in the figure is not identical to $g$ in \eqref{eq:shoving}, but differs by a normalizing factor.) 

Since the model is implemented as part of the normal \pythia~package, it can also be used to study subtle effects of introducing other scales to the collision. ATLAS has measured \cite{ATLAS:2019wzn} the ridge in collisions with a $Z$ boson present, instead of the normal QCD minimum bias collision systems. The presence of the $Z$ introduces the $m_Z$ scale as the largest hard scale of the collision, which could potentially alter the distribution (in $p_\perp$) of MPIs wrt. minimum bias. As shown in \figref{fig:shoving-results-pp} (right), the shoving model does a good job at reproducing the ridge also under these special circumstances.

Small collision systems are, however, not the main avenue for flow measurements. The shoving model has been applied, through Angantyr, to \AA collisions as well\footnote{As a historical note, it should be mentioned that the string shoving model is not the first attempt to generate anisotropic flow from string interactions. Abramovsky \textit{et al.} introduced a simple but similar idea already in 1988 \cite{Abramovsky:1988zh}, \emph{before} the idea of flow had been coupled to QGP formation at all.}. Here flow is quantified in flow coefficients ($v_n$'s), see e.g. ref. \cite{ALICE:2022wpn} for an overview. Currently the model does not perform well in full collisions, due to the presence of many soft gluons -- leading to many strings with many kinks -- which are difficult to handle technically \cite{Bierlich:2020naj}. Instead, toy approaches have been considered with some success.

\begin{figure}
	\includegraphics[width=0.5\textwidth]{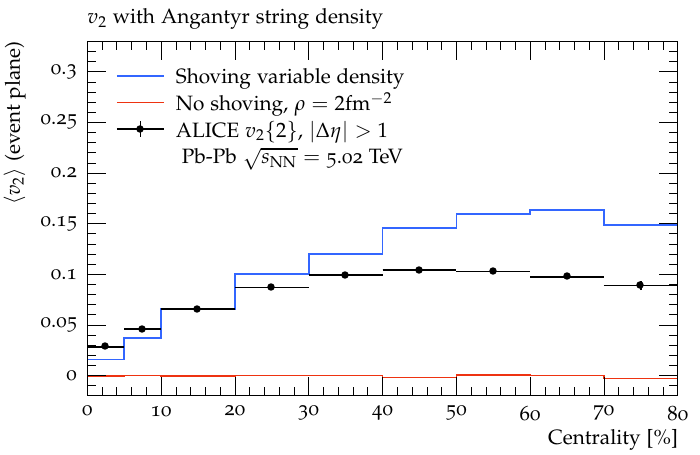}
	\includegraphics[width=0.5\textwidth]{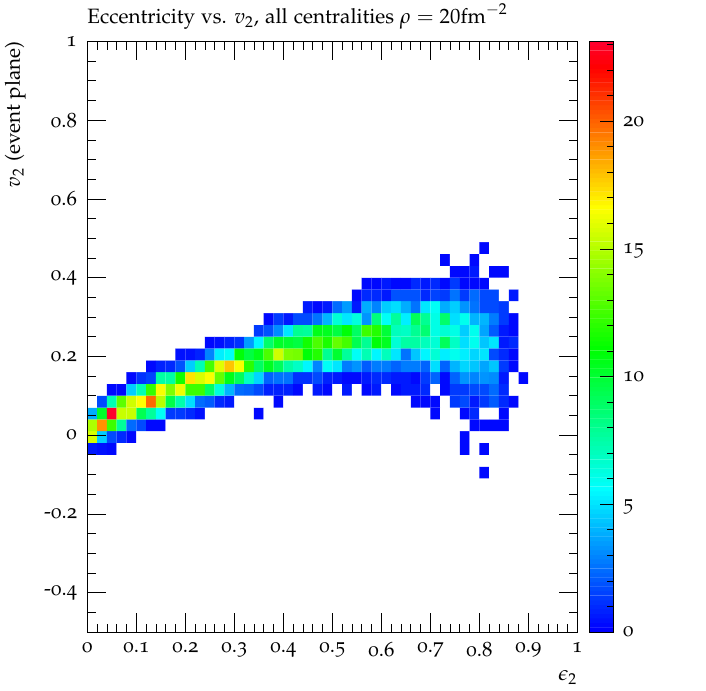}
\caption{\label{fig:shoving-results-aa}Shoving model results in \AA collisions, for toy initial conditions (see text). Comparison to $v_2$ in PbPb collisions at $\sqrt{s_{NN}} = 5.02$ TeV (left) shows a remarkable agreement with data, and studies of the correlation of $v_2$ vs. the eccentricity $\epsilon_2$ (right) shows a similar response as one would expect from hydrodynamic calculations. (Figures from ref. \cite{Bierlich:2020naj}, data from ALICE \cite{ALICE:2016ccg})}
\end{figure}

Since the multiplicity generated by a single string is well known (roughly one hadron per unit of rapidity), a system consisting of straight strings, i.e. without soft gluons, corresponding to the multiplicity of \AA collisions in a given centrality interval, can be set up. In ref. \cite{Bierlich:2020naj} this was done, along with spatial initial conditions resembling those of the given centrality interval. Furthermore, since the initial conditions are known in this case, flow does not have to be calculated in the same way as experiment\footnote{In experiment, the reaction plane angle is not known event-by-event, and one must construct flow coefficients from two -or multiparticle correlations. In calculations, in particular in \AA collisions where the non-flow contributions are small, the true reaction plane can be used.}, but can be calculated wrt. the true reaction plane, as well as correlations with the initial geometry. In \figref{fig:shoving-results-aa} (left), results for $v_2$ in this toy configuration, compared to data from ALICE \cite{ALICE:2016ccg}, is shown. While the agreement is by no means perfect, it should be emphasized that this calculation is performed using the same set of parameters as used for smaller collision systems in the same paper. This suggests that flow can indeed have the same origin across collision systems, and that the origin may be string interactions.

The normal explanation for flow in heavy ion collisions, is a hydrodynamic origin, due to the deconfined QGP phase. In hydrodynamic models \cite{Noronha-Hostler:2015dbi} as well as multiphase transport \cite{Wei:2018xpm}, it is known that the correlation between initial state eccentricity \cite{PHOBOS:2006dbo} and final state $v_2$ is linear. In \figref{fig:shoving-results-aa} (right), the same response is shown for the shoving model, revealing that it exhibits similar behaviour in dense systems. This suggests, as is perhaps not surprising, that hydrodynamic behaviour is \emph{not} limited to deconfined systems. From this follows that flow signals cannot be taken as proof of a QGP, indeed not if the same signal can be generated from the shoving model without QGP.

\subsubsection{Effects on jets}

As indicated above, the shoving model applies everywhere there are two or more strings available, irrespective of collision system or hard process. Since jet quenching in \AA collisions constitute some of the most solid evidence for QGP production, the -- so far unsuccessful -- search for jet modification in \pp (see e.g. \cite{Adolfsson:2020dhm}) is understandably a topic of some interest. When discussing jet modifications from string shoving, it is important to first emphasize that the model makes no distinction between jets, and the rest of the collision event. In the string model, every coloured parton must be connected to the rest of the event through a string, and when there are strings, there will be shoving effects. However, the small size of \pp collisions combined with the fast moving jet parton, makes jet modifications from string shoving very difficult to observe. However, string shoving is a final state interaction effect, and as such, jet modifications will show up given enough statistics/patience.

\begin{figure}
	\includegraphics[width=0.5\textwidth]{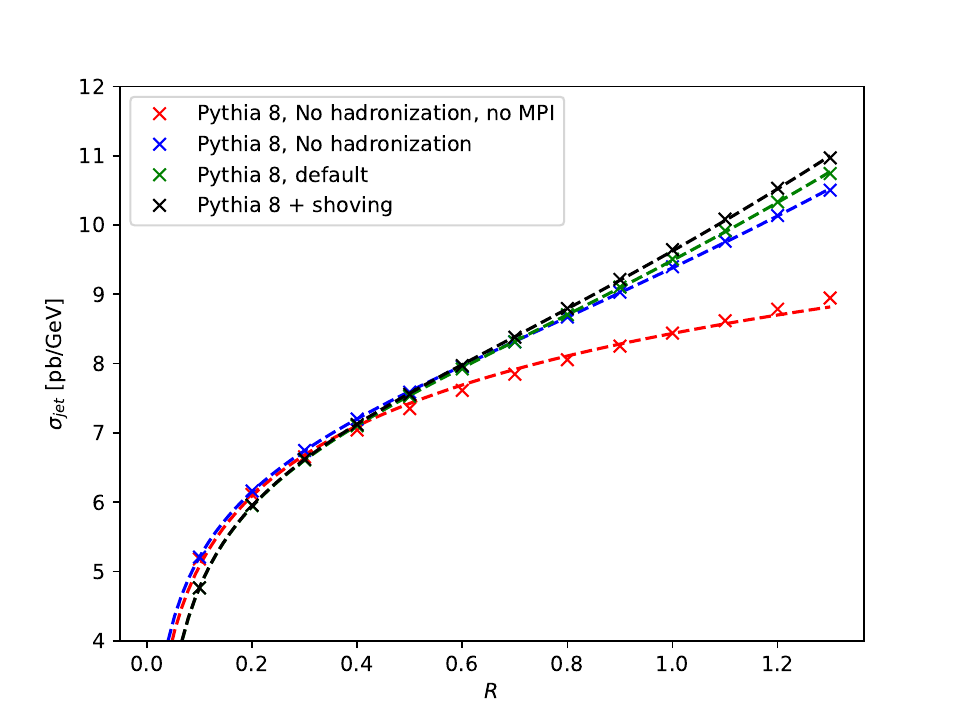}
	\includegraphics[width=0.5\textwidth]{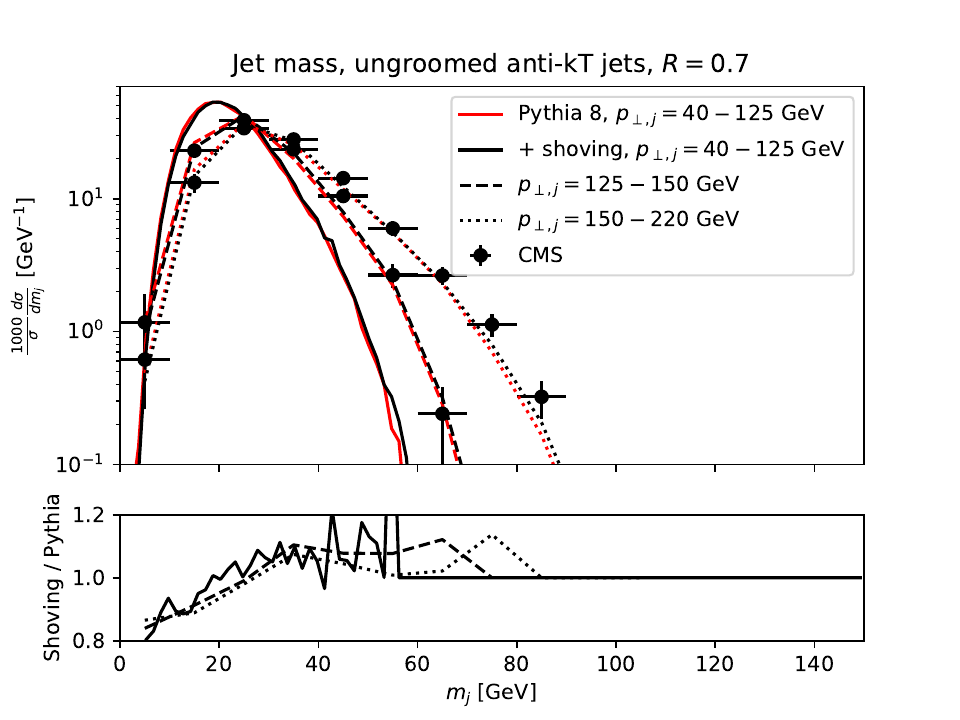}
\caption{\label{fig:shoving-results-jets}The effect on string shoving on jet cross section (details in text) vs. jet radius (left) shows a small effect, similar to that of hadronization. On observables such as jet mass (right) the effect can reach the 10-20\% level for small jet masses, out of reach of current experiments. (Figures from ref. \cite{Bierlich:2019ixq}, data from CMS \cite{CMS:2013kfv})}
\end{figure}

Consider a \pp collision where a $Z$ is produced together with a reasonably high $p_\perp$ jet. With a reasonable lower bound on jet $p_\perp$, here\footnote{As well as requiring a $Z$ with $p_{\perp,Z} > 40$ GeV, the jet is the leading anti-$k_\perp$ \cite{Cacciari:2008gp} jet with $\Delta \phi_{Z,j} > 3\pi/4$.} $p_{\perp,jet} > 80$ GeV, the jet cross section $\sigma_j = \int^\infty_{p_{\perp,0}} \mathrm{d}p_{\perp,j} \frac{\mathrm{d}\sigma}{\mathrm{d}p_{\perp,j}}$ is calculable in perturbative QCD, but receives large corrections from MPI and hadronization. In \figref{fig:shoving-results-jets} the jet cross section in $\sqrt{s} = 7$ TeV collisions is shown as function of jet $R$ ($= \sqrt{\Delta \eta^2 + \Delta \phi^2}$). As expected, the effect is not nearly as large as the effect from MPIs, which at large $R$ blows up, as the jet cone radius covers more and more of the underlying event. At large $R$ shoving contributes to the jet cross section at the same level as hadronization, i.e. at few percent level. The main point of the figure is twofold. First of old, one must be precise when asking for studies of jet modifications due to final state effects because, as shown, the jet definition can be doctored in such a way that it reaches way out in the underlying event, where even basic quantities like the jet cross section is affected. Second, that measurements of jet modifications due to so miniscule final state effects are indeed -- as expected -- difficult to measure. 

An example of a concrete observable, the jet mass of large $R = 0.7$ jets, is shown in \figref{fig:shoving-results-jets} (right), compared with measurements by CMS at $\sqrt{s} = 7$ TeV. Sizeable effects (over 10\%) show up at low jet masses, less than 20 GeV, but current measurements have no chance of distinguishing.

\subsection{Rope hadronization}
\label{sec:rope-hadronization}

After the shoving phase, string will hadronize. Hadrons will have properties as determined by equations (\ref{eq:frag-fun}) and (\ref{eq:mpt}). If the strings are, at this point, still overlapping, they may, however, form a ``rope'' \cite{Bierlich:2014xba}. Connecting back to the explanations on colour multiplets given in \secref{sec:colour-reconnection}, this is the effect that the string tension will be enhanced for higher multiplets\footnote{The rope hadronization idea dates back to the 1980s \cite{Biro:1984cf}, and has since been pursued in similar forms by many authors. Some very similar to the \pythia ropes, others in the same general class of string fusion models. Some implemented in event generators \cite{Merino:1991nq,Mohring:1992wm,Sorge:1992ej,Bleicher:2000us,Soff:2002bn,Amelin:1994mc,Armesto:1994yg}, and others with purely theoretical work \cite{Bialas:1984ye,Kerman:1985tj,Gyulassy:1985oqt,Braun:1991dg,Braun:1993xw}.}

$\mathbf{SU(3)}$ multiplets can conveniently be characterized by the two quantum numbers $(p,q)$, which can be given the heuristic interpretation as the number of parallel and anti-parallel strings acting together coherently when overlapping. The simple example of $\mathbf{3} \otimes \mathbf{3} = \mathbf{6} \oplus \mathbf{\bar{3}}$ can as such be written as $(1,0) \otimes (1,0) = (2,0) \oplus (0,1)$. The quadratic Casimir ($C_2$) of a given $(p,q)$ multiplet, can be written in terms of $p$ and $q$, normalized by $C_2$ of a triplet. Since the string tension scales like $C_2$ in lattice calculations \cite{Bali:2000un}, this can be used to calculate the effective string tension ($\tilde{\kappa}$) of any multiplet as:
\begin{equation}
	\frac{\tilde{\kappa}(p,q)}{\kappa(1,0)} = \frac{C_2(p,q)}{C_2(1,0)} = \frac{1}{4}(p^2 + pq + q^3 + 3p + 3q). 
\end{equation}
In the \pythia rope model, the rope hadronizes by breaking up one ``thread'' at the time. One such breaking will therefore take the multiplet from the state $(p,q) \rightarrow (p-1,q)$. The effective string tension \emph{in the string break} therefore becomes:
\begin{equation}
	\label{eq:enh-single}
	\frac{\tilde{\kappa}}{\kappa} = \frac{2p + q + 2}{4}.
\end{equation}

\begin{figure}
\includegraphics[width=0.8\textwidth]{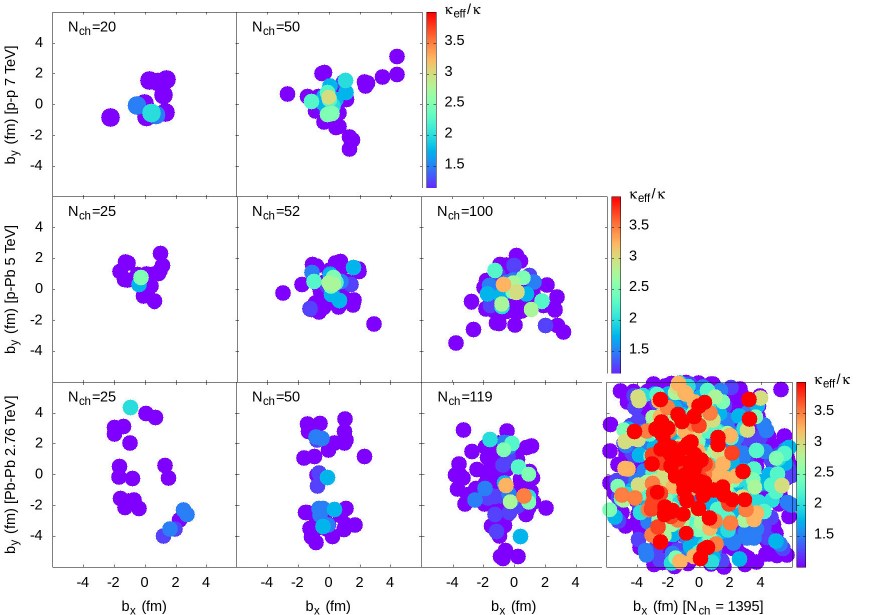}
\caption{\label{fig:energy-density}Production points of hadrons in impact parameter space for representative collisions of \pp at 7 TeV (top row), pPb at 5.02 TeV (middle row) and PbPb at 2.76 TeV (bottom row). The hadron production points are coloured according to the rope tension, giving an image of the energy density reached in the ropes. (Figure from ref. \cite{Bierlich:2022ned}.)}
\end{figure}

The energy density in a rope can quickly grow large enough to make a dramatic impact on final state observables (see \secref{sec:rope-observables}). It will, however, not exceed more than around a factor 2 above the vacuum string tension, except in very central \AA collisions, where a factor 2 more can be reached. In QGP based models it is often argued (see e.g. ref. \cite{Chen:2015wia}) that initial state energy densities ($\mathrm{d}E/\mathrm{d}^3x$), which for example in CGC based models can reach over 100 GeV/fm$^3$ in PbPb collisions at LHC \cite{Schenke:2012hg,Schenke:2012wb,Bierlich:2020naj}, are so large that plasma creation is inevitable. Crucially the energy density in a rope scenario never reaches such levels, as shown in \figref{fig:energy-density}. Other models of string interactions, crucially the string perculation model \cite{Bautista:2019mts}, has the opposite conclusion: that the percolation of strings is potentially the mechanism responsible for QGP formation \cite{Ramirez:2020vne}.

\subsubsection{Effects on inclusive production}
\label{sec:rope-observables}

When the string tension rises, the main effect is that suppression of strangeness in \eqref{eq:mpt} decreases. The derived quantity $\rho$ is replaced by an effective quantity $\tilde{\rho}$, which can be obtained directly as:
\begin{equation}
	\tilde{\rho} = \exp\left(-\frac{\pi(m^2_s - m^2_u)}{\tilde{\kappa}}\right) = \rho^{\kappa/\tilde{\kappa}},
\end{equation} 
where the exponent can be obtained directly from $(p,q)$ through \eqref{eq:enh-single}. Similar expressions exist for the other parameters of the hadronization model, see refs. \cite{Bierlich:2014xba,Bierlich:2022ned}.

\begin{figure}
\includegraphics[width=0.8\textwidth]{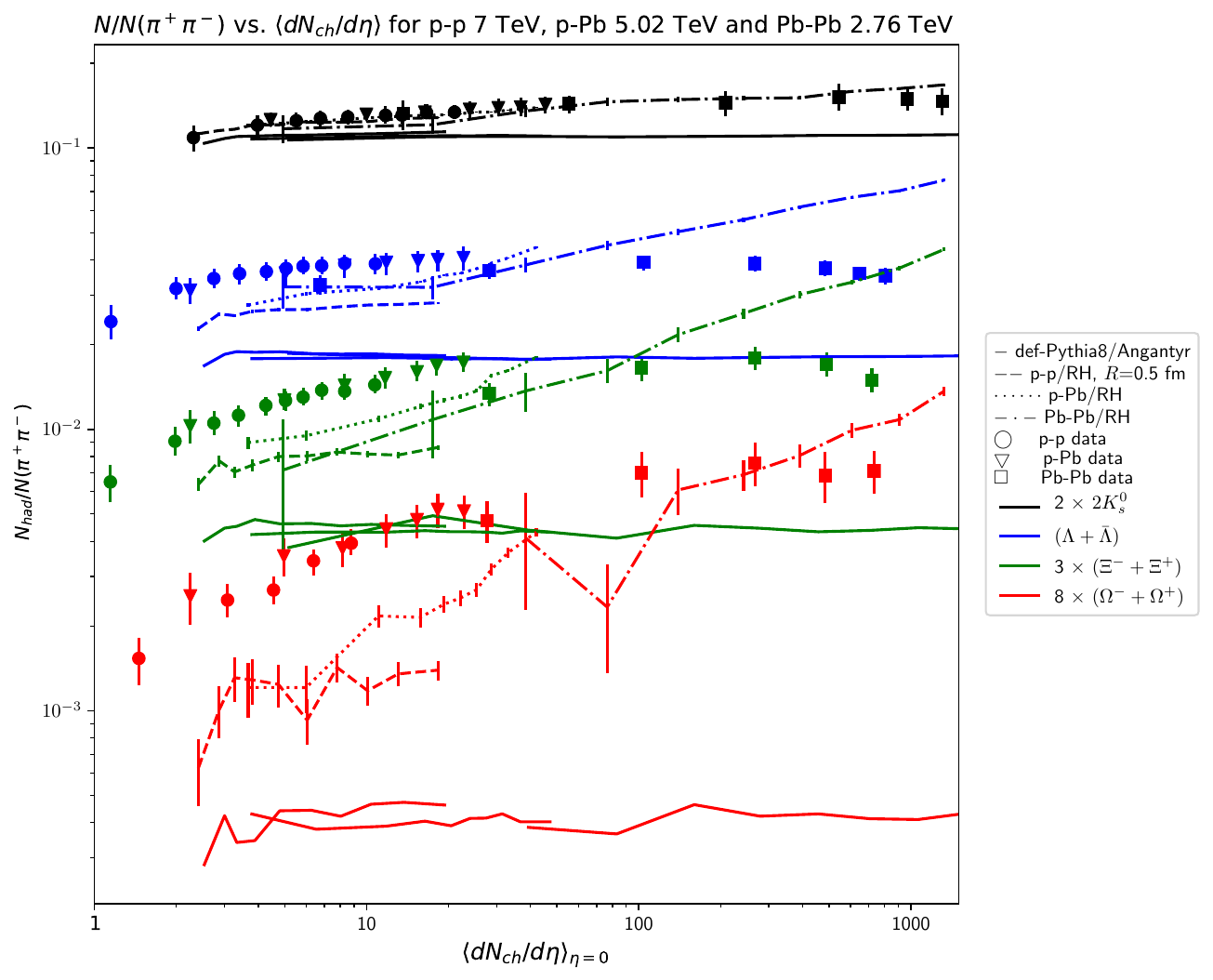}
\caption{\label{fig:rope-all}Production of multi-strange hadrons with rope hadronization in pp, pPb and PbPb. (Figure from ref. \cite{Bierlich:2022ned}. Data from ALICE \cite{ALICE:2016fzo}.)}
\end{figure}

Crucially, the base value of $\rho$ is not touched, but fixed to \epp data once and for all. Furthermore, the model does not require any fitting to heavy ion data at all. Once the geometry of the \AA collision is set up, everything else is fixed by smaller collision systems.

Production of multi-strange baryons is now affected both by the introduction of junction in the colour reconnection mechanism explained above, as well as the enhancement of strange quark production. In \figref{fig:rope-all} a summary of strange multi-strange hadron production across pp, pPb and PbPb is shown. While the model does not fully capture the data, it exhibits the correct trend across all collision systems. As was the case for string shoving, work on extending this facet of the model to all collision systems is still an ongoing endeavour.

\section{Conclusion and outlook}
\label{sec:conclusion}

The discovery of collective effects in small collision systems has spurred renewed interest in hadronization models, a field that has otherwise been quite dormant since the late 1980s and early 1990s. This mini-review discusses ongoing attempts to construct a QGP-free model to describe behaviour in both small and large collision systems, traditionally ascribed to QGP formation. However, developments are also occurring in other areas, both connected and unrelated to these endeavours. For the string hadronization model, this includes attempts to modify the usual \eqref{eq:mpt} to a thermal one \cite{Fischer:2016zzs}, colour reconnection corrections to the top quark mass \cite{Argyropoulos:2014zoa}, calculations of spatial vertices in the fragmentation \cite{Ferreres-Sole:2018vgo} underlying hadronic rescattering, hyperfine splitting effects \cite{Bierlich:2022vdf}, and more theoretical attempts to investigate the single string entanglement entropy \cite{Berges:2017zws}, with some phenomenological considerations also made \cite{Hunt-Smith:2020lul}. Similarly, for the \herwig cluster model, efforts such as cluster reconnections \cite{Gieseke:2017clv,Duncan:2018gfk} have made progress in the production of multi-strange baryons, akin to string interactions.

A very recently emerging direction involves the formulation of hadronization models using Machine Learning-based empirical models, with a special focus on Deep Learning techniques. Complementary approaches are being developed for both the string and the cluster models. The HadML initiative has demonstrated proofs-of-principle that the cluster model can be parametrized using GANs \cite{Ghosh:2022zdz}, and how such models can be tuned to real data \cite{Chan:2023ume}, providing a path forward for real applications. The MLHad initiative \cite{Ilten:2022jfm} has similarly shown how the string model can be parametrized using normalizing flows \cite{Bierlich:2023zzd}, as well as how string hadronization can be re-parametrized \cite{Bierlich:2023fmh} to allow for more efficient parameter variation and ultimately more robust comparisons to data. While it is still unclear whether such efforts will contribute directly to a deeper physics understanding of hadronization, it is evident that they will aid in making better quantitative comparisons to data, and ultimately in ruling out models that cannot globally describe data.

Ahead of us lies the high luminosity phase of the LHC. This phase is expected to bring new insights into hadronization, not least through improved statistics on heavy flavour production, particularly in the baryon sector. This advancement is expected to impact the colour reconnection models as new knowledge emerges. The upcoming electron-ion collider, as well as further reanalysis of LEP data until the FCC-ee, will hopefully shed further light on collectivity in the smallest systems.

\section*{Acknowledgements}
I wish to thank Manuel Szewc for a well placed comment on Deep Learning techniques.
Support from Vetenskapsrådet, contract number  2023-04316 and the Knut and Alice Wallenberg foundation, contract number 2017.003, is gratefully acknowledged.

\bibliographystyle{utphys}
\bibliography{refs.bib}

\end{document}